\documentclass[runningheads]{cl2emult}

\usepackage{makeidx}  
\usepackage{graphicx} 
\usepackage{subeqnar} 
\usepackage{multicol} 
\usepackage{cropmark} 
\usepackage{eso}      
\makeindex            

\begin{document}
\title*{Terapixel Surveys for Cosmic Shear}
\toctitle{----}
%
%
\titlerunning{Terapixel Surveys for Cosmic Shear}
%
\author{Y. Mellier\inst{1,2}
\and L. van Waerbeke\inst{1,3}
\and M. Radovich\inst{4,1}
\and E. Bertin\inst{1,2}
\and M. Dantel-Fort\inst{2}
\and J.-C. Cuillandre\inst{5}
\and H. Mc Cracken\inst{6}
\and O. Le F\`evre\inst{6}
\and P. Didelon\inst{7}
\and B. Morin\inst{7}
\and R. Maoli\inst{8,1,2}
\and T. Erben\inst{9}
\and F. Bernardeau\inst{10}
\and P. Schneider\inst{11}
\and B. Fort\inst{1}
\and B. Jain\inst{12}
}
\authorrunning{Mellier et al.}
%
%
\institute{Institut d'Astrophysique de Paris, 98 bis Bd Arago, 75014
Paris, France
\and Obs. de Paris, DEMIRM, 77 av. Denfert Rochereau, 75014 Paris,
France
\and CITA, 60 St. George Street, Toronto M5S 3H8, Canada
\and OAC, via Moiariello 80131 Napoli, Italy
\and CFHT, P.O. Box 1597, Kamuela Hawaii 96743, USA
\and LAS, 13376 Marseille Cedex 12, France
\and SAp, CE Saclay, 91191 Gif-sur-Yvette Cedex, France
\and Dipartimento di Fisica, Universit\`a di Roma ``La Sapienza'', Italy
\and MPA, Karl-Schwarzscild Str. 1, 85748 Garching, Germany
\and SPhT, CE Saclay, 91191 Gif-sur-Yvette Cedex, France
\and Universitaet Bonn, Auf dem Huegel 71, 53121 Bonn, Germany
\and John Hopkins University, Dept. of Physics, Baltimore MD21218, USA
     }

\maketitle              

\begin{abstract}
The recent detections of cosmic shear signal announced by several groups have  
 demonstrated the feasibility of this challenging program and
  convinced  astronomers of its potential for cosmology.   
 Cosmic shear analysis demands to handle Gigabytes of data in order 
  to probe several square degrees in subarcsecond deep 
imaging mode.  The success of these
  surveys is  sensitive to the designs of the 
observation strategy,  the organization of 
  the data reduction pipelines and the links of the data base 
with surveys like X-ray or spectroscopic follow up.  
 We describe the cosmic shear surveys we have carried out at the VLT and
at CFHT and the way we handle this huge data set in a more general
context including the VIRMOS and the XMM-LSS surveys, and the future 
CMB surveys.
\end{abstract}

\section{Cosmology with weak lensing}
Large-scale structures of the universe  induce gravitational 
lensing effects which accumulate on  
  photons emitted by distane sources. On deep CCD images,   
 they revealed themselves as a weak modification of the shape of
 galaxies which adds to their  intrinsic ellipticity
   to produce a cosmological  shear signal, called 
  the {\sl cosmic shear}. The light beam deformation  
   witnesses the history of mass density fluctuations 
  from the emitting (lensed) sources to the observer.  Therefore,   
    it is a signature 
 of the cosmological scenario of structure formation  and its 
  study should provide   
    interesting clues on several cosmological quantities, 
like the cosmological parameters, the
 power spectrum of density fluctuations and the biasing\footnote{http://www.iap.fr/LaboEtActivites/ThemesRecherche/Lentilles/LentillesTop.html}. \\
The cosmological potential of cosmic shear has been 
 pointed out by a decade of theoretical studies.  
Since most of the condensations crossed by photons 
  are extended large-scale structures,  their cumulative 
lensing effects can be computed by applying the 
perturbation theory to low mass density contrast lenses.  
In the  case of a single lens plane and assuming  the
shape of the power spectrum of density fluctuations
 is a power law (ie $P(k) \propto k^{n}$),  perturbation theory
applied to weak cosmological lensing provides interesting 
insights on  the  sensitivity of 
 the gravitational convergence, $\kappa$ (ie the projected 
 mass density of lenses), and the gravitational 
 shear, $\gamma$ (ie the distortion), to 
cosmological models:
\begin{itemize}
\item For small perturbations, the variance of the convergence
averaged over an angular scale $\theta$, $\langle \kappa(\theta)^2
\rangle$ depends on cosmological quantities in a simple way:
\begin{equation}
\langle \kappa(\theta)^2 \rangle^{1/2} \approx \ 0.01 \ \sigma_8
\ \Omega_m^{\ 0.75} \ z_s
^{\ 0.8}
\left({\theta \over 1^o}\right)^{-(n+2)/3} \ ,
\end{equation}
 where $\sigma_8$ is the normalization of the power spectrum, $z_s$ the
redshift of sources.
\item Likewise, the skewness of the convergence on angular scale
$\theta$,
$s_3(\theta)$, writes:
\begin{equation}
s_3(\theta) \approx \ 40 \ \Omega_m^{\ -0.8} \ z_s^{\ -1.35} \  
\end{equation}
 (Bernardeau et al \cite{bernardeauetal}).
 Hence, when they are used jointly,  
 the variance and the skewness 
   can constrain simultaneously $\Omega_m$ and $\sigma_8$.
\item The gravitational convergence can be easily related to
 the gravitational shear, $\gamma$:
\begin{equation}
\langle \kappa(\theta)^2 \rangle=\langle \gamma(\theta)^2 \rangle \ .
\end{equation}
Since in the weak lensing regime $\gamma$ is measured directly from the
gravity-induced ellipticity of galaxies, the cosmic shear 
  can be estimated almost directly from the measurement of galaxy 
  ellipticities.
\item  The amplitude of the weak lensing signal is not beyond the 
reach of present-day instruments.  Jain \& Seljak \cite{js} 
or van Waerbeke et al \cite{vwetal2} (2000b)   explored  the
non-linear
regime of mass density fluctuations which mostly changes the 
 convergence on small scales. 
 The non-linear evolution of the power spectrum increases the
amplitude
of the  cosmic shear  by a factor of two as compared  
  to the linear prediction. Hence, on angular scales below 10', 
 the cosmic shear is already measurable with  
  current  ground-based telescopes.
\end{itemize}
\section{Definition of the cosmic shear survey}\label{section2}
Cosmological distortion only increases the ellipticity of lensed 
  galaxies by  a few percents. Its detection, which is hampered   
  by artificial distortions of similar amplitude,
 can only be recovered statistically from the morphological study of   
 thousands of galaxies
 spread over several degrees of the sky.  Hence, in order to recover 
  the cosmic shear signal it is necessary to carry out 
  a deep wide field survey and to handle a huge amount of data.  \\
Van Waerbeke et al \cite{vwetal99} used 
  extensive simulations in order to design the survey
 and to infer its minimum angular coverage to recover  
  cosmological quantities.  It turns out that a shallow survey covering
 a large field of view is a better strategy than a  deep 
cone.  An optimal design seems to be a  survey 
  covering 10$\times$ 10 deg$^2$ up
to $I=24$. At this depth, the redshift distribution of the sources 
can be  constrained from photometric and spectroscopic
redshifts with enough accuracy  to separate most realistic cosmological models
 with a good significance level. This is  the 
strategy we are preparing for the MEGACAM survey.\\
We can also use these simulations in order to design a more modest 
cosmic shear survey feasible on a short time scale.  The
    results  listed in Table \ref{sizesurvey} 
  show that  the variance of $\kappa$ can be 
  measured with a good significance if the  survey size 
  covers at least one deg$^2$,  whereas 
   about 10 deg$^2$ are needed to estimate its skewness. This can
already be done with present-day instrumentation, like the UH8K, the
CFH12K or WFI and CFHT or at ESO.  

\begin{table}[t]
\caption{\label{sizesurvey}Expected signal-to-noise ratio on the
measurement of
the variance and the skewness of the convergence for two extreme
realistic
  cosmological models.  In the first
column, the size of the field of view (FOV) is given. The
signal-to-noise
 ratio is computed from the simulations done by
van Waerbeke et al  (1999). The top line of this table describes
shortly some details of the analysis.  The redshift of the sources and
the galaxy number density correspond to a typical 2-hours exposure on a
4-meter telescope.}
\begin{center}
\begin{tabular}{|l|c|c||c|c|}
\hline
\multicolumn{5}{|c|}{$z_s=1$, Top Hat
Filter , $n=30$ gal.arcmin$^{-2}$} \\ \hline
\hline
  FOV &
\multicolumn{2}{|c|}{ {S/N Variance}} &
\multicolumn{2}{|c|}{ {S/N Skewness}} \\
\cline {2-5}
 (deg.$\times$deg.) &  $\Omega_m=1$  &
 {$\Omega_m=0.3$}  &
 {$\Omega_m=1$}  &
 {$\Omega_m=0.3$}  \\ \hline
 1.25$\times$1.25  &  7  &  5 &  1.7 &   2 \\ \hline
 2.5$\times$2.5  & 11  & 10 & 2.9 &  4 \\ \hline
 5$\times$5  & 20  & 20 & 5 &  8\\ \hline
 10$\times$10  & 35 & 42 & 8 &  17\\ \hline
\end{tabular}
\end{center}
\end{table}

\begin{table}
\caption{\label{summarysurvey}Summary of the 5 cosmic shear surveys 
which announced a detection during the first semester of 2000. The CFHT
data were obtained with the UH8K and CFH12K CCD cameras. The R and I 
limiting magnitudes give  a reasonnable estimate of the redshift of
the sources, which should be around one. }
\begin{center}
\begin{tabular}{|l|c|c|c|c|}
\hline
Reference & Telescope & Lim. Mag. & FOV & Nb. fields \\
 \hline
van Waerbeke et al \cite{vwmeetal} & CFHT & I=24 & 1.7 deg$^2$ & 5 \\
Wittman et al \cite{wittetal} & CTIO & R=26 & 1.5 deg$^2$ & 3 \\
Bacon et al \cite{baconetal1}& WHT & R=24 & 0.5 deg$^2$ & 13 \\
Kaiser et al \cite{kaiseretal}  & CFHT & I=24 & 1.0 deg$^2$ & 6 \\
Maoli et al  \cite{mvwmetal}& VLT-UT1 & I=24 & 0.5 deg$^2$ & 45 \\
\hline
\end{tabular}
\end{center}
\end{table}

\section{Detection and analysis of first cosmic shear signals}\label{section3}
\begin{figure}
\centering
\includegraphics[width=1.1\textwidth]{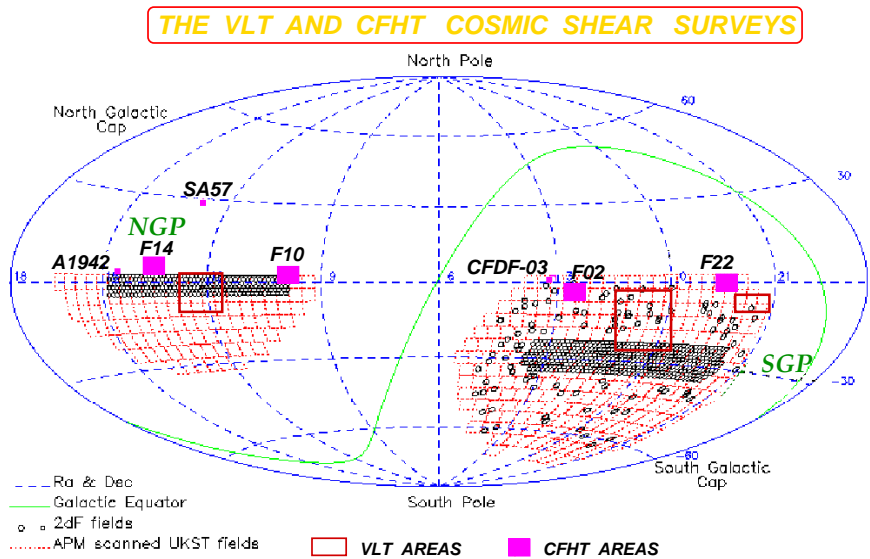}
\caption[]{Positions of the areas covered by the CFHT and VLT cosmic shear
surveys. The filled areas are the targets covered at CFHT.  The open
rectangles
 delineate the areas inside which the 45 VLT targets were selected.
\label{oursurvey}}
\end{figure}

\begin{figure}
\centering
\includegraphics[width=1.0\textwidth]{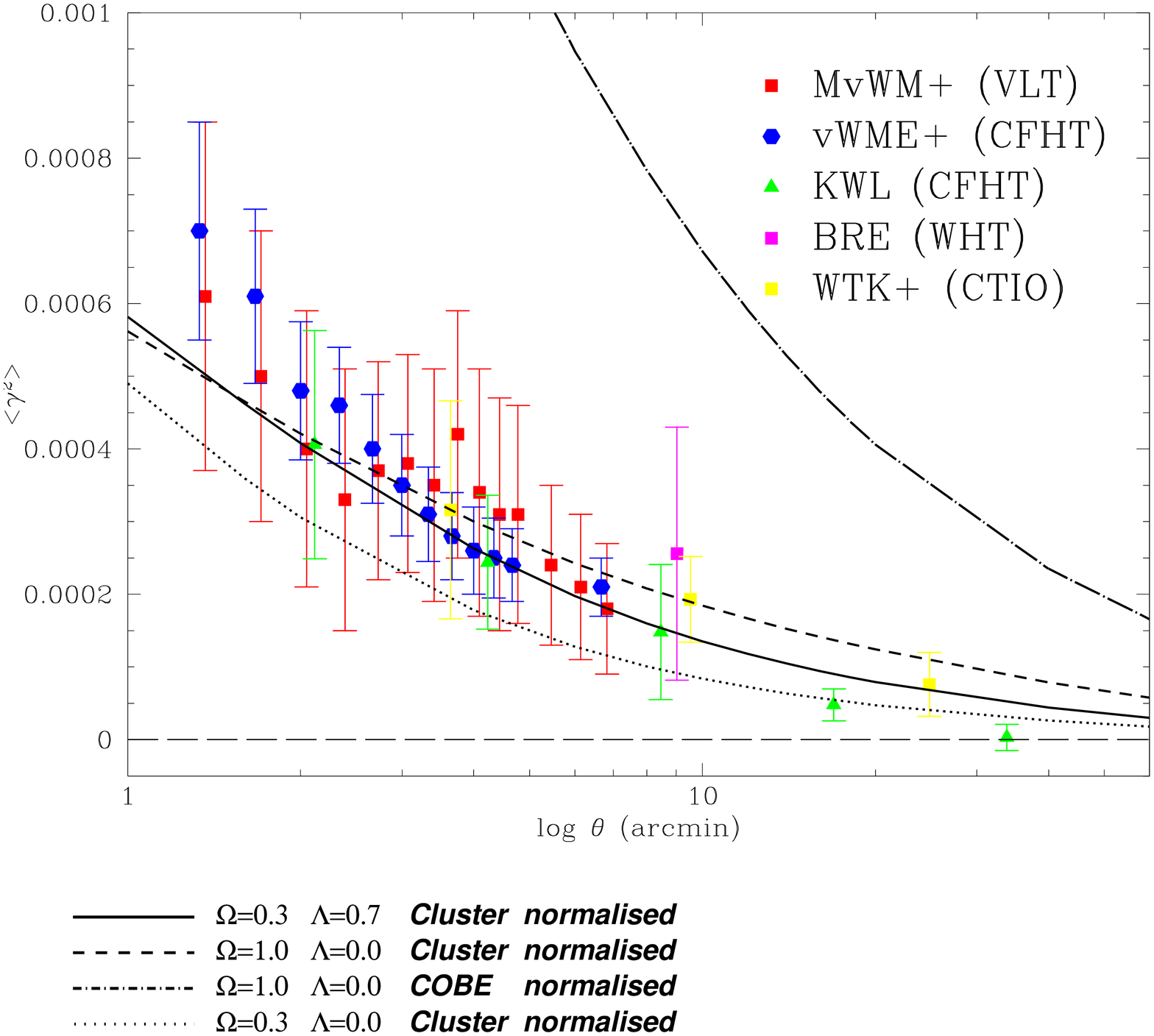}
\caption[]{
The recent results of cosmic shear measurements. The  works
 referred to as Maoli et al 2000 (MvWM+),
   van Waerbeke et al 2000 (vWME+),
 Kaiser et al 2000 (KWL), Bacon et al 2000 (BRE) and Wittman et al
 2000 (WTK+).
Some predictions of cosmological models are also plotted, assuming
sources at $z_{eff}=1$ and using the non-linear evolution of
power spectrum according to the coefficients given by Peacock (1999).
}
\label{cosmicshear}
\end{figure}

The four teams which carried out a cosmic shear  survey 
 used different instruments, observed different fields
of view and used different
 techniques to analyze the data and correct for the PSF anisotropy.
 The  CFHT and VLT surveys  reported in van Waerbeke et al
 \cite{vwmeetal} and Maoli et al 
 \cite{mvwmetal} respectively (see Fig. \ref{oursurvey})
 consist in two  independent
  data sets, which  enable us to cross-check
our results and to explore the reliability of our
 corrections of  systematics.  The 45 VLT are of special interest
because the data were obtained in service mode which permits 
  to get an homogeneous sample of data obtained in very similar
 depth and seeing
 conditions. The VLT targets are spread over more than 1000 squares degrees, 
  each
of them being separated from the others by at least 5 degrees.  These
uncorrelated fields provide a direct measurement of the cosmic variance,
without need of simulations. \\
The results of the four surveys are summarized  in 
Table \ref{summarysurvey} and in Fig. \ref{cosmicshear}.  
  The most striking
 feature on this plot is
  the remarkable similarity of the results in the range 1' to 10' .
This is a very strong point which validates the
  detection and guarantees that they are reliable and  robust, despite
   concerns about systematics.
%
%
The comparison of the measurements with some typical cosmological
models displayed in Fig. \ref{cosmicshear} (the non-linear evolution of
the power spectrum is computed with the coefficient given in Peacock
\cite{pik}, which seems to provide lower amplitudes of the
 variance of the shear
 than previous coefficients) leads to the following
conclusions:
\begin{itemize}
\item The simultaneous use of independent data provided by the five groups
   permits to rule out  
 some models with a very high significance.  In particular,
  the SCDM COBE-normalized model is rejected to at least a 5-$\sigma$
level.
\item In contrast, most popular cluster-normalized models
 fit  the data reasonably well and the discrimination between 
  them is not possible. This illustrates
that error bars are still too large and also that the variance of the
shear
  is not enough to break the degeneracy ($\sigma_8,\Omega_m$).  
 Only once the skewness of the convergence will be measured 
 we will be in much better position to constrain cosmological
  scenarios.
\end{itemize}
The depth of these surveys corresponds to sources at redshift of about
$ z \approx 0.8-1$. The typical efficiency function, which describes the
lensing strength of the lenses as function of the redshift
distributions of the lenses and the sources,
  should therefore peak at redshift $z \approx 0.4$.
 On angular scales between 1' and 10', since the non-linear structures
dominate
  the signal, most of the cosmic shear is produced by structures having
physical sizes of about $0.2 - 1.0 $ $h^{-1}$ Mpc.  Hence, 
 these cosmic shear surveys
   mainly probe  weak cosmological lensing produced by
 clusters of galaxies and compact groups.  \\
The constraints provided by cosmic shear are formally similar to
those from cluster abundances obtained from
  counts of clusters in optical or X-ray surveys (Eke et al \cite{Eke96},
 Eke et al \cite{Eke98},
 Bridle et al \cite{Bridle99}). Depending on the angular scales,
 the variance of the cosmological convergence writes:
\begin{equation}
\langle \gamma^2\left(\theta\right) \rangle^{0.5} \propto
 \sigma_8^{^{\ \approx 1.1}} \Omega_m^{^{\ \approx 0.7}}  \ \ , 
\end{equation}
whereas, for  cluster abundances the constraints have formally the
following
 dependences:
\begin{equation}
\sigma_{_8}  \Omega_m^{^{\ \approx 0.55}} \approx 0.6  \ .
\end{equation}
The cosmic shear has the advantage of being a direct measurement of the
lensing
effects produced by dark matter.  In contrast, the cluster abundances
measures the fraction of massive clusters from the light
distribution, which implies, either empirical relation
between light and mass (like emissivity-temperature relation), or
  assumptions
  of the geometry and the physical state of the baryonic and
non-baryonic components. \\

\section{Massive data processing for cosmic shear surveys}\label{section4}
The estimations of the survey size done by van Waerbeke et al (1999) 
  provide the minimum angular size needed to measure both the variance
 and the skewness of the convergence.  However,
 we ultimately plan to produce a projected mass map of the sky 
 and  to deproject the power spectrum of the mass 
density fluctuations over physical sizes as large as 100 Mpc.  We 
therefore need to probe at least 100 square-degrees and to   
 get informations on the redshifts of the lenses and the sources.\\
The CFHT cosmic shear survey we are currently carrying out covers 
  the sky area of  
   the VIRMOS imaging survey  which will
  cover  an angular size of 16 square degrees in BVRI at CFHT, plus 
  the U-band and a fraction in J and K at the ESO telescopes. In total, more 
than 256 CFH12K pointings in four colors are expected. Including the
calibration files, it corresponds to about 3 Terabytes of data and more than
 40, 000 FITS files.   This huge amount of data prefigures the 
 complexity of data handling in future surveys, like those that will be done 
at CFHT with the 20K$\times$18K MEGACAM camera \cite{bouladeetal}.  \\
The TERAPIX data center installed at IAP is designed in order to 
provide facilities to MEGACAM users but is already widely used to 
process UH8K and CFH12K images.  Its role is to develop softwares and 
pipelines and to provide hardware and technical assistance to  astronomers 
  who are using wide field cameras like UH8K, CFH12K, WFI and, later,
  MEGACAM and OMEGACAM.   
Many softwares commonly used by observers or in other pipelines 
  are actually developed, updated or validated at the TERAPIX data center  
 (This is for instance the case of SExtractor, WEIGHTwatcher or 
FLIPS which is currently developed at CFHT).  TERAPIX is also developing 
  new softwares/pipelines/interfaces for astrometric corrections, co-addition 
of images, new image display and preprocessing tools, like
  the PANORAPIX image display,  (see Radovich et al' poster in these proceedings). In 
  addition, it is 
preparing a new Object Oriented Data Base which will drive and will organize
 the future MEGACAM survey in the perspective of the enlarged 
  MEGACAM/VIRMOS/XMM survey.\\
Regarding the hardware side, the TERAPIX data center has 3 COMPAQ EV5,
  EV6 and EV67 XP1000 computers,
4 LINUX PCs, 3 X-terminals,  more than 2 Tb of RAID-5 disk space and 
  usual DDS3/DLT4000/DLT7000 tape drives.  This is not the final 
configuration of the data center, but its size matches well the 
  needs for WFI, UH8K and CFH12K images.  \\
TERAPIX has processed 1.7bytes the VIRMOS/WL 
of images from December 1999 to August 2000, corresponding to more than
 22, 000 FITS files. This survey turns out to be very 
helpful to prepare the pipelines for MEGACAM and to foresee the 
  potential problems we could face on in the future.  
At the end of the survey, more than  
  4 Terapixels will have been processed by TERAPIX.  \\
One important issue
 is the organization of the data base which in principle should 
  control the pipeline and keep track of the complete history of the 
processing and of the archiving.  Since our weak lensing surveys 
should be coupled with the VIRMOS redshift surveys\footnote{http://www.astrsp-mrs.fr/virmos/}, the XMM 
 large scale structure survey\footnote{http://vela.astro.ulg.ac.be/themes/spatial/xmm/LSS/}, as well as the 
  VLA and SZ follow up, we expect to provide  a data base
 which will be easily handled by the consortium.   These wishes
 have not yet any concrete impact on TERAPIX. At this stage we are 
still trying to provide some specifications on the basis of the 
scientific objectives we have in mind, as those we summarize in the 
next section, which will certainly demand multi wavelength data bases.

\section{Cosmic shear in a virtual observatory context}\label{section5}
The detection of a cosmic shear signal is  a first step toward a
comprehensive investigation of the dark matter of the universe and its
role on formation and evolution of structures and galaxies.  Since 
 a complete understanding of the cosmological scenario implies  
   both the baryonic and non-baryonic  contents of the universe, cosmic
shear data only is not enough and the survey should be completed by 
   additional data from optical/NIR, X-ray  and radio surveys or 
even CMB surveys.  The 
issues addressed below summarize some important goals which  
 need  multiple data sets coming from several surveys 
and concerning data of different nature.
\begin{itemize}
\item {\sl Redshift distribution of the sources:} as shown by Eq. (1)
and
 (2), both the variance and the skewness of the convergence depend 
 on the source redshift.  We therefore do need   
spectroscopic follow up and calibration of photometric redshifts 
  of the galaxies used in the cosmic shear sample.
\item {\sl The source clustering problem:} 
Due to galaxy clustering, the amplitude of the gravitational shear
may strongly vary from one line-of-sight to another.
 The average redshift distribution of the sources can 
be biased by the galaxies located within the massive structure,
which affect the value of the convergence in a similar way.
The recent simulations done  by Hamana et al
\cite{hamana} and (Thion et al \cite{thion})
   show that source clustering 
   significantly perturb the signal by 20\%.
  A simple way to avoid this problem is to  reduce the redshift 
range of the selected sources.  Hamana et al \cite{hamana}
  demonstrated that the uncertainty can be reduced to
 1\% if one uses sources within a redshift range of $\Delta \ z \approx
0.2$. So, in principle deep surveys like the imaging+spectroscopic 
  VIRMOS will enable us to solve this issue.
\item {\sl Test on the linear biasing:} Schneider \cite{sch1998} and 
van Waerbeke \cite{vw2000} (2000c) pointed out that the cross correlation 
of galaxy distribution with the aperture mass statistics only depends on the 
cosmological models and the linear biasing factor.  When the cross
correlations on two different angular scales are compared, one can 
 probe the evolution of the linear biasing with angular scale and 
with redshift. This estimator, which may be of crucial importance to 
constrain the scenario of galaxy formation, is insensitive to 
cosmological parameters which makes this tool very attractive. 
   In practice, it means that one need to couple 
  multiple data sets: photometric, redshift  and shear catalogues 
  averaged on various angular/physical scales.
\item {\sl Baryonic vs. non-baryonic matter distribution:} part of the cosmic 
shear surveys we are carrying out maps the dark matter distribution in the 
sky area which will be also covered by the XMM large scale structures
  survey.  The common area will provide 
  simultaneously, the stellar light, the hot-gas and the dark matter 
distributions on scales ranging from 1 arc-minutes to one degree.  
  At least a complete sample of clusters and groups of galaxies will be
 investigated in detail in order to recover the baryonic to non-baryonic
 fraction of matter, the
 mass profiles of each distribution in clusters and groups of galaxies
  and the large-scale filamentary distributions of both components.
\\
Alternatively, the XMM X-ray selected sample of clusters of galaxies 
  will provide a fairly accurate estimate of the cluster 
abundances up to $z \approx 1$. From this, one can infer 
  the normalization of the power spectrum and therefore break the 
      degeneracy between $\sigma_8$ and 
  $\Omega$ which is expressed in Eq. (1).
\item {\sl Relation between Large-scale structures and AGNs/EROs:}
 as for clusters and groups, the XMM surveys will probe the large-scale
distributions of X-ray point sources emitted by AGNs. The most 
recent Chandra and XMM deep surveys seem to show that a significant 
fraction of EROs are indeed X-ray sources as well. Cosmic shear, 
  X-ray and deep photometric catalogues 
  can therefore be used jointly to explore the relation 
 between AGNs, EROs and large-scale structures
   detected by weak lensing mass reconstruction.
\item {\sl Cross correlating galaxy and CMB weak lensing signal:} 
 likewise galaxies, the temperature fluctuations of the CMB map can be
distorted by foreground lenses along the line of sights.  In principle,
 the distortion pattern of the CMB map does contain similar informations
 as galaxies, but with the advantage that the redshift of the source is
 well known (ie $z = 1000$!).  Bernardeau \cite{bern1998} 
 explored this idea but concluded that the weak lensing signal 
produced on the CMB will be marginal.  A better strategy is to 
  analyze the gravitational shear simultaneously on both the 
CMB temperature maps and the galaxies.  
 Van Waerbeke et al \cite{lud2000} (2000d) pointed out that the 
correlation of these signals will significantly improve the 
detection of lensing on  CMB maps produced by the Planck Surveyor
mission. 
\item {\sl Coupling real data set with mock catalogues:} this is 
 an important point which should not be neglected in a survey. 
 The real data set must be compared to mock catalogues illustrating 
  realistic universes and analyzed in exactly the same conditions as the
 real data. This enables to estimate accurately the cosmic variance and
 the sources of systematics.  Mining the sky does necessarily imply to
  make mock catalogues available.
\item {\sl Intrinsic correlated polarization of galaxies}
If the intrinsic orientations of galaxies
 are not randomly distributed, their
coherent alignment may correlate to the geometry of large
scale structures in which they are embedded. If so, the
coherent alignment produced by weak lensing will be corrupted by the intrinsic
 alignment of the galaxies and a mass reconstruction
 based on the shear pattern will be strongly contaminated by spurious
weak
lensing signal.  Recent analyses carried out by Croft \& Metzler
\cite{croft} and Heavens et al. \cite{heavens}  conclude
that on scales smaller than 10 arcminutes the intrinsic correlation
should not contaminate the weak lensing signal, provided the survey
 is deep enough in order to probe distant lensed galaxies.  In 
 contrast for shallow survey the conclusions are more 
pessimistic and the intrinsic correlation could even dominate 
the signal.  \\
The intrinsic 
ellipticity problem is an issue that can be addressed by using different 
  surveys.  For instance, the nearby galaxy sample provided by the 
SDSS can easily check whether such correlations exist and 
  up to which angular scales it dominates the cosmic shear signal.
\end{itemize}
  
\section{Conclusions}
Thanks to the recent detections of cosmic shear signal, we know that 
 weak lensing surveys can now provide reliable informations on the 
  large-scale dark matter distribution in the universe which 
  would be inaccessible otherwise.  The scientific impact 
  of these surveys should increase rapidly.  On going  
 wide field cosmic shear surveys are going to produce the first measurements
 of the variance,  the skewness of the convergence in less than 
  one year and we expect to infer
  the properties of the  power spectrum of mass density fluctuations
  and the linear biasing up
  to degree scales within the next 5 years.   \\
These exciting perspectives contrast with the worrysome technical 
 issues we may face on regarding data handling. Besides the 
 hundreds of Terabytes of data which have to be processed, we 
 also have to think about archiving and data mining.  
The optimal use of weak lensing statistics demands to handle 
  simultaneously the baryonic and non baryonic content of the universe. 
  Optical/NIR and X-ray/SZ surveys dealing with the baryon content 
  and its
evolution with look back time must be analyzed together with the 
dark matter distribution and interpreted  in  cosmological contexts 
which can be described by numerical simulations. The complexity 
  of the joint data analyses is certainly a challenge for the 
  future.  The solution we are preparing for MEGACAM is a joint 
 multi-wavelength survey which is designed in advance in order to 
optimize the strategy and the archiving.  In the future, we hope that
  the  MEGACAM/VIRMOS/XMM/VLA program will provide an 
  easy-to-use and homogeneous database  which will 
include for the first time the dark matter content for multipurposes 
  cosmological projects.

\section*{Acknowledgements} We thank M. Pierre, 
  T. Hamana and A. Thion for fruitful
discussions. YM thanks the organizers of the meeting 
  for travel support. This work was supported by the TMR 
  Network ``Gravitational
Lensing: New Constraints on Cosmology and the Distribution of Dark
Matter'' of the EC under contract No. ERBFMRX-CT97-0172.


\clearpage
\addcontentsline{toc}{section}{Index}
\flushbottom
\printindex


\begin{thebibliography}{99}
\addcontentsline{toc}{section}{References}

\bibitem{bernardeauetal}Bernardeau, F., van Waerbeke, L., Mellier, Y.
1999, A\&A 322, 1.

\bibitem{js}Jain, B., Seljak, U. 1997, ApJ 484, 560.

\bibitem{vwetal2} van Waerbeke, L., Hamana, T., Scoccimaro, R., Colombi,
S., Bernardeau, F. 2000, submitted (2000b).

\bibitem{vwetal99}van WaerbeKe, L., Bernardeau, F., Mellier, Y. 1999,
A\&A 342, 15.

\bibitem{vwmeetal} van Waerbeke, L., Mellier, Y., Erben, T. et al 2000,
A\&A 358, 30.

\bibitem{mvwmetal} Maoli, R., Mellier, Y., Van Waerbeke, L. et al 2000,
Astro-ph/0008179.

\bibitem{wittetal}Wittman, D.M., Tyson, A.J.,, Kirkman, D. et al 2000,
Nature 405, 143..

\bibitem{baconetal1}Bacon, D., R\'efr\'egier, A., Ellis, R.S. 2000,
 Astro-ph/0003008.

\bibitem{kaiseretal}Kaiser, N., Wilson, G., Luppino, G. 2000,
Astro-ph/0003338.

\bibitem{pik} Peacock, J.A. 1999. ``Cosmological Physics". Cambridge.

\bibitem{Eke96} Eke, V.R., Cole, S., Frenk, C.S. 1996, ApJ 282, 263.

\bibitem{Eke98} Eke, V.R., Cole, S., Frenk, C.S., Henry, P.J. 1998, ApJ
298, 1145.

\bibitem{Bridle99}Bridle, S.L., Eke, V.R., Lahav, O. et al 1999, MNRAS
310, 565.

\bibitem{bouladeetal} Boulade, O., Charlot, X., Abbon, P. et al 2000. 
 SPIE's "Astronomical Telescopes and Instrumentation 2000", Munich, 2000 .

\bibitem{hamana} Hamana, T., Colombi, S., Mellier, Y.  2000, Proceedings
of the XX$^{th}$ Rencontres de Moriond. "Cosmological Physical with
Gravitational Lensing". Y. Mellier, J.-P. Kneib, M. Moniez, J.
Tran Thanh Van eds. Astro-ph/0009459.

\bibitem{thion} Thion, A., Mellier, Y., Bernardeau, F., Bertin, E.,
Erben,
T., van Waerbeke, L. 2000, Proceedings
of the XX$^{th}$ Rencontres de Moriond. "Cosmological Physical with
Gravitational Lensing". Y. Mellier, J.-P. Kneib, M. Moniez, J.
Tran Thanh Van eds. Astro-ph/0008180.

\bibitem{sch1998} Schneider, P. 1998. ApJ 498, 43.

\bibitem{vw2000} van Waerbeke, L. 2000c, A\&A 334, 1.

\bibitem{bern1998} Bernardeau, F, 1998. A\&A 338, 767.

\bibitem{lud2000} Van Waerbeke, L.; Bernardeau, F.; Benabed, K. 2000d,
  ApJ, 540, 14.

\bibitem{croft} Croft, R.A.C., Metzler, C. 2000, Astro-ph/0005384.

\bibitem{heavens} Heavens, A., R\'efr\'egier, A., Heymans, C. 2000,
Astro-ph/0005269.

\end{thebibliography}
\end{document}